\documentclass[reprint,preprint
amsmath,
amssymb,
aps,
pra,
floatfix,
]{revtex4-2}

\usepackage{graphicx}
\usepackage{dcolumn}
\usepackage{bm}
\usepackage{amsmath}

\begin{document}


\title{High-order harmonic generation  probing of a ring-opening reaction}

\author{Lauren Bauerle}
    \altaffiliation[Also at ]{Department of Chemistry, University of Colorado}
\author{Agnieszka A. Jaron}%
    \altaffiliation[Also at ]{Department of Physics, University of Colorado}
\affiliation{JILA, University of Colorado, Boulder, CO-80309, USA}%

\date{\today}

\begin{abstract}
Using Time-Dependent Density Functional Theory (TDDFT)  nonlinear nonperturbative response of the molecular system is studied for photoisomerization reaction. The 1,3-cyclohexadiene photoisomerization is probed by the high-harmonic generation (HHG) process induced by the ultrashort few femtosecond intense laser pulse. For each of the long-lived or stable isomers, we present results for HHG spectra and HHG ellipticities, as well as multi-photon ionization. Moreover, the laser-induced electronic excitations are calculated and the effect of excitations on HHG spectra is discussed. We describe the characteristic properties of the nonlinear response for each of the isomers that can be used for ultrafast detection of the isomers by tracking the specific changes in HHG response.

\end{abstract}


\maketitle

\section{Introduction}
Ultrafast intense laser pulses can be a powerful tool for controlling atomic and molecular dynamics. Often the manipulation of the electron dynamics is achieved by controlling the properties of the molecular wavepackets (for review on this topic see e.g. \cite{review1} and references therein). In the case of probing performed by ultrashort laser pulses one of the advantages is that one does not have to consider the effect of the modifications of the potential energy surface (PES) topology by the electric field of the laser pulse. This can be useful for studies of chemical reactions since the dynamics driven by the laser-modified multidimensional PES can differ from laser-free dynamics, which anyway can be quite complex, particularly near conical intersections. Ultrashort pulses allow for focusing studies on the ultrafast electron dynamics without adding the complexity of modification of the nuclear dynamics and consideration of laser-induced changes of the reaction even for high-intensity laser pulses. In particular, few femtosecond FWHM laser pulses allow for studies of ultrafast effects related directly to electron rearrangement during chemical reactions \cite{Li2010}, detailed studies of electron excitations \cite{Woerner2017}, and monitoring of ionization \cite{Kupper2021, Li2010, Garg2024}, all while the modification of the PES during laser-molecule interaction can be neglected due to the difference in the timescales between the dynamics of electrons and nuclei. 
Alternatively, strong laser field pulses can also be applied in such a way as to purposefully modify the dynamics by opening new reaction channels that are otherwise inaccessible. For example, one such topic that has been studied more extensively is coherent control of non-adiabatic dynamics at avoided crossings and conical intersections (See e.g. \cite{Humeniuk_2016}).

The experimental studies of the dynamics often involve pump and probe laser pulses, where the pump laser pulse initiates the electronic transition and in this way prepares the molecular wavepacket, while the probe laser pulse is designed to "investigate" the properties of the evolving molecular wavepacket prepared by the probe pulse. Pump probe spectroscopy studies using time-delayed ultrafast probing to follow the progress of chemical reactions often mention "molecular movies" in this context and the result of the visualization of the results of ultrashort probe pulse interaction with the system at a given time delay is called a snapshot. 

\begin{figure}[h!]
    \centering    \includegraphics[width=1\columnwidth]{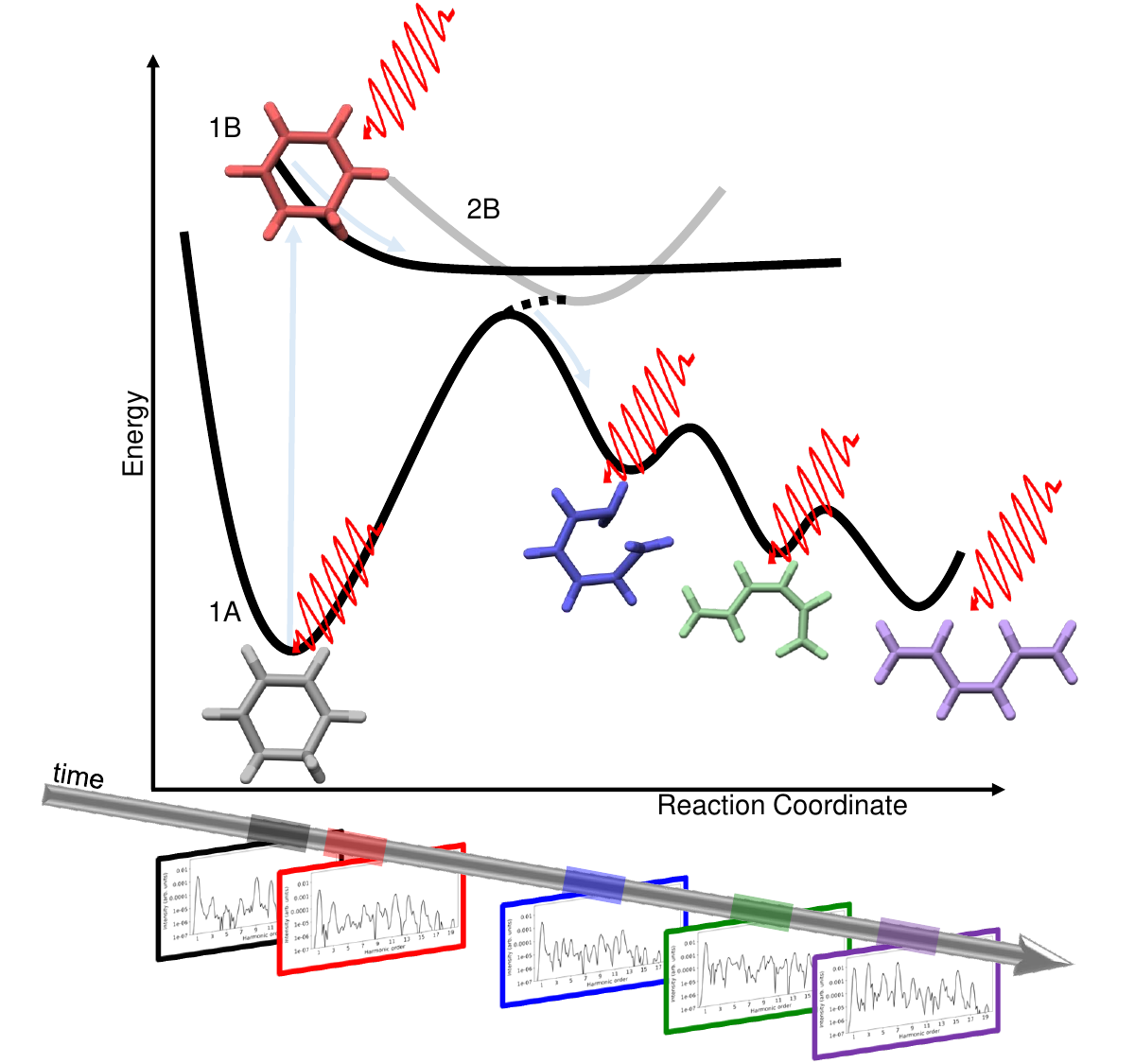}
    \caption{Scheme of the HHG probing of the photoisomerization of 13CHD with illustration of the geometries of the long-lived isomers and the product of the reaction.}
    \label{fig:scheme}
\end{figure}

The present study is devoted to theoretical calculations towards making a molecular "movie" of the progress of the photoinduced ring-opening reaction of 1,3-cyclohexadiene (13CHD) that produces 1,3,5-(Z)-hexatriene (135HT). This reaction is an example of an electrocyclic chemical reaction that is mediated by one or more conical intersections \cite{trushin1997, wolf2019}.
Interest in this reaction is driven by the role played in the photobiological synthesis of vitamin D$_3 $\cite{jacobs1979_2,havinga1961} but the photoisomerization of 13CHD is much simpler and less computationally demanding. It has been studied extensively both theoretically and experimentally (See e.g. \cite{kuthirummal2006,trushin1997,karashima2021} and references therein). 
The reaction begins with photoexcitation from the ground state of 13CHD  to the bright $1B$ state, i.e. an electron is promoted from the Highest Occupied Molecular Orbital (HOMO)  to the Lowest Unoccupied Molecular Orbital (LUMO)   by a UV laser pulse. The excited state molecular wavepacket then rapidly crosses onto the $2A$ state where it continues to evolve to the $1A$ ground state. At the $2A/1A$ conical intersection, the wavepacket bifurcates and approximately half of the population returns to the 13CHD  and the other half proceeds along the remaining ring-opening reaction path to form 135HT \cite{wolf2019}. The process is represented schematically with relevant isomers in Figure \ref{fig:scheme}. This isomerization reaction can proceed differently if a different laser pulse is used for the pump stage. 

The reaction is probed by a time-delayed ultrashort intense laser pulse that drives the High Order Harmonic Generation (HHG) process and the ultrafast HHG radiation properties directly "record the snapshot" of the reaction.  HHG process is one of the nonlinear intense field-driven processes that require theoretical description without the support of the perturbation theory or linear response theory. The same reasons why one cannot apply linear response approximation cause difficulties with numerical simulations and pose challenges for obtaining correct reliable and physically relevant theoretical predictions. In the illustration of analysis of the HHG process, one often utilizes a simplified physical picture that translates the results into 
 a sequence of simpler mechanisms that correspond to mechanisms discovered according to the semi-classical description of the HHG from atoms, called the three-step model   \cite{corkum1993, lewenstein1994}.
This model assumes the HHG proceeds through 3 steps: ionization, followed by "rescattering" which means propagation of the free or nearly free electron interacting with the electric field of an intense laser pulse, and then followed by 3rd and last process, namely recombination. Within numerical simulation of the propagation of the atomic/molecular wavepacket, one does not separate these mechanisms. 

In the next section, we present a review of the theoretical methods used to obtain our results and computational details of the numerical simulations of the HHG process.

\section{Theory and computational details}
We present results of numerical simulations in which Time-dependent Density Functional Theory (TDDFT) and Density Functional Theory  (DFT) calculations were performed using the open-source software Octopus \cite{octopus}. TDDFT/DFT is based on mapping interacting system properties into properties of the noninteracting electron system with a specially designed external potential called Kohn - Sham potential. One can learn about properties of the interacting systems, even often in the regime where other methods are computationally intractable, because of the advantage that one numerically solves time-dependent single electron Kohn-Sham equations, 
\begin{equation}\label{tddft}
    \epsilon_{i} \phi_{i}(\textbf{r},t) = \left[ -\frac{1}{2} \nabla^{2} + V_{KS}(\textbf{r},t)  \right] \phi_{i}(\textbf{r},t),
\end{equation}
for each valence orbital $\phi_{i}(\textbf{r},t)$ and the Kohn-Sham potential $V_{KS}$, is defined as, 
\begin{equation}
V_{KS}({\bf r}, t) = 
V_{ext}({\bf r},t) + 
\int \frac{\rho({\bf r}', t)}{|{\bf r}-{\bf r}'|} d{\bf r}' + 
V_{xc}({\bf r})
\end{equation}
and includes the external potential due to the interaction of the electron with the $N$ nuclei in the molecule and with the time-dependent electric field: 
\begin{equation}
V_{ext}({\bf r},t) = 
\sum\limits_{i=1}^{N} -\frac{Z_i}{|{\bf R}_i-{\bf r}|} + 
E_0(t) \sin(\omega t) \sum\limits_{k=1}^{n} {\bf r}_k \cdot {\hat\epsilon}
\end{equation}
where $Z_i$ is the charge of the $i$th nucleus, $\hat\epsilon$ is the laser polarization direction, $\omega$ and $E_0(t)$ are the angular frequency and the time-dependent amplitude of the laser field.  The electron density is usually defined as, 
\begin{align}
\rho({\bf r}, t) = \sum\limits_{k=1}^{N} f_k |\phi_k({\bf r}, t)|^2,
\end{align}
where $N$ denotes occupied valence Kohn Sham orbitals, $f_k$ is the electron population in the $k$-th Kohn-Sham orbital $\phi_k({\bf r}, t)$.

The photoisomerization reaction of interest has recently been studied experimentally using ultrashort intense laser pulses and high harmonic generation process  \cite{kaneshima2018}. The pump-probe character of the experimental study permits us to focus on HHG response for more stable isomers that correspond to minima in the reaction path.
In the direct vicinity of the conical intersections, where the changes of the molecular geometry happen on faster timescales, one might expect that there are lower chances of conditions favorable for production of the coherent HHG radiation. Consequently, we study ionization and HHG properties for five isomers  1,3-cyclohexadiene (13CHD), excited 1,3-cyclohexadiene (13CHD*) in 1B state, cZc-1,3,5-hexatriene (cZcHT), cZt-1,3,5-hexatriene (cZtHT), and (3Z)-1,3,5-hexatriene (135HT). 

\begin{figure}[h!]
    \centering
    \includegraphics[width=1\columnwidth]{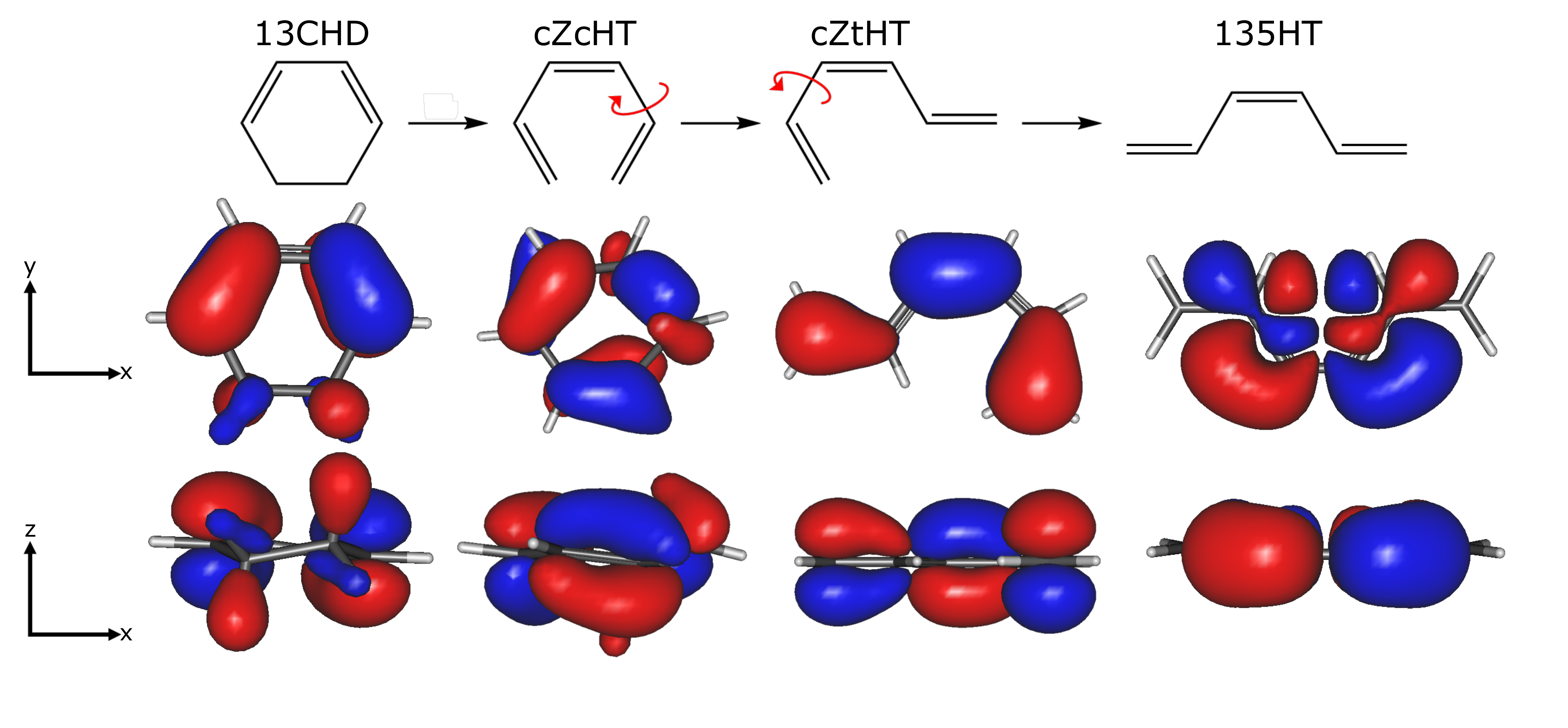}
    \caption{Photoisomerization reaction represented  schematically (top row) and the corresponding 3D isosurface visualization views (middle and bottom rows) of HOMO for each isomer, from left to right: 13CHD,  cZcHT, cZtHT, 135HT}
    \label{fig:rxn}
\end{figure}

Since the systems considered here have 14 atoms, all electrons calculations without further approximations are very time-consuming even within the TDDFT method. We use the local density approximation (LDA)  for the exchange-correlation energy. 
Let us note that it has been observed that TDDFT with LDA approximation can overestimate ionization yields, however, any functional that allows for the correct description of the asymptotic potential requires a prohibitively long simulation time for the present case. Ionization potentials for many molecules are lower than ionization potentials for atoms, which means for example that for given laser pulse parameters one can generate fewer higher order harmonic and the overall shape of the HHG spectrum can have a downward slope toward the end of the spectrum at the cut-off. Our calculations are performed for rather moderately high intensity that does not drive substantial ionization so we do not expect to have the results strongly affected by this. 
Moreover for each isomer in our calculations, the 32 valence electrons are active in the simulations, while the core electrons are approximated by the pseudopotentials \cite{Troullier_1991}. 

The photoisomerization reaction has been observed in the experiment to occur over hundreds of femtoseconds \cite{kaneshima2018}. The probe pulse used in the present calculations has a FWHM of approximately 11 fs. Consequently, it is expected that assuming a fixed position of the nuclei during the interaction with the laser pulse is a reasonable approximation. However we have performed preliminary calculations with moving ions for systems studied in this article and there are no discernible differences between frozen geometry and moving ions calculations regarding the results presented in present paper. 

The photoisomerization reaction has been observed in the experiment to occur over hundreds of femtoseconds \cite{kaneshima2018}. The probe pulse used in the present calculations has a FWHM of approximately 11 fs. Consequently, it is expected that assuming a fixed position of the nuclei during the interaction with the laser pulse is a reasonable approximation. However, we have performed preliminary calculations with moving ions for systems studied in this article and there are no discernible differences between frozen geometry and moving ions calculations regarding the results presented in the present paper. 

The photoisomerization reaction is assumed in both theory and experiment to be initiated by a pump pulse of wavelength 267 nm \cite{wolf2019,kaneshima2018}, and the ultrafast electron dynamics is probed by 800 nm intense laser pulses \cite{kaneshima2018}. We use the same laser pulse wavelength for the probe pulse as has been used in experiment \cite{kaneshima2018}. 
Let us note that the 800 nm wavelength does not correspond to any important one-photon transition for the isomers investigated in the paper. 

In order to be able to estimate the effects of the excitations during interaction with the intense laser field we have calculated the absorption spectra using the Casida method, which is a Linear Response Time Dependent DFT (LR-TD-DFT) method and involves diagonalizing a matrix in the basis of occupied-unoccupied orbital pairs \cite{Casida1996, Casida1998} and in present case the implementation uses the same molecular Kohn Sham orbitals as the initial ground state wavefunctions that are used for time-dependent propagation method. The results of these calculations can be seen as superimposed red lines in HHG spectra presented in Figure \ref{fig:hhg_x} and we discuss these results below in more detail. 

Please note that due to the highly demanding nature of the computations requiring parallel multi-thread high memory execution and the properties of the queuing system of the supercomputer we had access to, we were forced to consider modified laser pulse parameters with respect to experiments performed in \cite{kaneshima2018}. 
In our simulations, probe laser pulses are assumed to have a peak intensity of 3.7$\times$10$^{13}$W/cm$^{2}$, which is lower than the probe pulse used in the experimental setup  \cite{kaneshima2018}. A trapezoidal pulse envelope was used to allow for more simulation time at the maximum intensity of the laser pulse and to limit the effects on HHG due to the ultrashort nature of the assumed probe pulses. The trapezoid pulse shape has a ramp-up of 3 fs (1.12 optical cycles), a plateau of 8 fs (3 optical cycles), and a ramp-down of 3 fs (1.12 optical cycles). 

The laser pulse is assumed to be polarized along the $\hat{x}$-axis, which is parallel to the molecular, $x$-$y$ plane. This laser pulse polarization is expected to produce higher harmonic yields since for this direction the HOMO is aligned predominantly along the laser pulse polarization direction. The relative orientation of molecules and laser pulse polarization direction can be seen in Figure \ref{fig:rxn}. Based on conclusions of orientation-dependent ionization of di- and polyatomic molecules (see for example \cite{AJ1,AJ2}), one expects strong contributions from HOMO and smaller contributions from other valence orbitals. 
Let us note that ionization is often considered  the "source of electrons" and  the first step in the commonly used simple man model of the HHG process that has provided intuitive interpretations in the past
(see e.g. \cite{Miller2016, Ciappina2007, Ciappina2008}).

Moreover, for other relative orientations of the molecule and laser field polarization we expect to have weaker HHG signals due to for example the nodal plane's involvement and the related destructive interference's effects on the HHG signal \cite{Alharbi_2017}. In addition, the pump stage selects molecules that are oriented in a way that maximizes the probability of excitation that initiates the reaction. The randomly oriented non-excited molecules might contribute signal to the total HHG signal that can be considered as a background noise and since we focus here on changes that can be seen over the progress of the reaction the rotational averaging is less important for that purpose. 
In this paper, we report our results of numerical simulations corresponding to this complex experiment. We focus on analysis of the direct results of the simulations of microscopic HHG response and refrain from additional attempts for example to perform focal averaging or rotational averaging. 

Let us note that in the past we have considered a macroscopic response description of HHG from the nitrogen molecular ion and we could observe how only the contributions corresponding to the part of the laser pulse with the highest intensity are most important for the properties of the coherent signal  \cite{Joyce2020}. 
This means in our present case the effective "interaction time" is shorter than 8 fs (corresponding to the flat part of the laser pulse envelope) which in addition motivates frozen nuclear geometry calculations.

The time-dependent Kohn-Sham equations are propagated using the Crank-Nicolson propagator \cite{crank_nicolson_1947} with a time-step of 0.02 a.u.. The calculations proceed on a spatial grid that expands in the $\hat{x}$- and $\hat{y}$-directions from -40 to 40 a.u. In the $\hat{z}$-direction, the grid goes from -30 to 30 a.u. The grid is smaller in $z$ since one expects the majority of electron dynamics to be driven in the $x$-$y$ plane since only a few atoms in any of the isomers are located off of the $x$-$y$ plane and the laser pulse is polarized along the $\hat{x}$-axis. The interaction with the $\hat{x}$-polarized laser pulse is expected to have a rather small effect on dynamics in the $\hat{z}$-directions, allowing for the box to be smaller in that direction.
The grid has spacing in each direction of 0.3 a.u.. For the outermost 5 a.u. of the box a complex absorbing potential, $V_{abs}(\textbf{r})$, is used. The complex absorbing potential is added to the Kohn-Sham potential, and the total effective potential can be written as,
\begin{equation}\label{cap}
    V_{eff}(\textbf{r},t) = V_{KS}(\textbf{r},t) + iV_{abs}(\textbf{r}),
\end{equation}
where $V_{abs}(\textbf{r})$ is zero in the inner region of the simulation box and rises smoothly until the edge of the box \cite{marques2003},
\begin{equation}\label{cap2}
    V_{abs}(\textbf{r})=
    \begin{cases}
    0 , & \textbf{r} < \textbf{r}_s \\
    -i\theta\sin^2\left(\frac{\pi(\textbf{r}-\textbf{r}_s)}{2L}\right), & \textbf{r}\geq \textbf{r}_s
    \end{cases}
\end{equation}
where $\theta$ is the height and $L$ is the width of the absorbing potential.  
A complex absorbing potential is designed to remove the effects of reflection from the edges of the simulation box by absorbing any part of the wavepacket that makes it to the edges. The portion of the wavepacket that is absorbed is analyzed to study the total ionization since in this type of simulations the  “lost” electron density i.e. the absorbed portion of the wavepacket describes the ionized wavepacket. 

Initial geometries for 13CHD and 135HT are assumed to be according to  The Computational Chemistry Comparison and Benchmark Database \cite{cccbdb} and further optimized using Gaussian 16 \cite{g16} using the B3LYP/6-31G(d) basis set \cite{g16, basisset}. For 13CHD*, the geometry and orientation are assumed to be the same with only the orbital occupations changing. The geometries and orientation for intermediate isomers cZcHT and cZtHT are taken from the supplementary information of \cite{wolf2019}. 

Within our simulations, the time-dependent dipole moment is used to calculate the high harmonic spectrum. A Blackman window of the form $
    W(t) = 0.42 - 0.5 \cos(2\pi t/M) + 0.08 \cos(4 \pi t /M)$, 
where $M$ is the number of points in the dipole and zero padding are applied to the time-dependent dipole before calculating the Fourier transform in order to get the harmonic spectrum. 

The dominant dipole for the HHG spectrum is the dipole in the $\hat{x}$-direction, the direction of the laser pulse polarization. A majority part of the harmonics radiation is expected to be in the direction of laser pulse polarization, but in this case, it is possible to have a significant enough dipole in either of the perpendicular directions ($\hat{y}$ and $\hat{z}$). This means the harmonics generated are elliptically polarized. Let us note that in the past TDDFT has provided very good agreement with experimental results for the ellipticity of HHG generated from nitrogen molecule \cite{xia_2014}, even without focal averaging. For atoms and systems with inversion symmetry, only odd harmonics are expected.
When molecules are considered the presence of even harmonics is possible  \cite{Kopold_1998, Nalda_2004}. 

In the next section, we present the analysis of the results of the numerical simulations.

\section{Results}
\subsection{HHG spectra}
In this section, we present results related to the molecular HHG process. One of the characteristics of the HHG spectra is the cut-off frequency. 
The primary cut-off for the harmonics is calculated using the formula 
$I_p + 3.17 U_p$, where $I_p$ is the approximate adiabatic ionization potential and $U_p$ is the ponderomotive energy. The formula sometimes used to approximate the vertical ionization potential $I_{p}^{vertical} = - \varepsilon_{HOMO}$ is underestimating the ionization potentials in our case. To have better approximations of the ionization potentials, we calculate them using the difference between the total energy of the neutral molecule and molecular cation calculated with DFT B3LYP exchange-correlation functional \cite{g16}. One can only calculate the vertical $I_{p}$ for cZcHT and cZtHT due to the geometry optimization required for calculations of the adiabatic $I_p$. Results of the vertical ionization potentials and the corresponding HHG spectra cut-offs in Table \ref{tab:important_energies}.

The adiabatic ionization potentials calculated by us 
are for 13CHD 8.023 eV and for 135HT 8.267 eV which are not very different from the vertical ionization potentials calculated. 
The vertical ionization potentials calculated by us are closer to the experimentally measured ionization energies, namely for 13CHD 8.25 eV and for 135HT 8.30 eV  \cite{bieri1977}. The cut-off calculated using the vertical $I_p$ for the HOMO-1 is shown as well, because the HOMO-1 energy is within two photons, making it possible to have ionization and contributions to HHG from HOMO-1 in addition to the HOMO. Ionization from that orbital and subsequent recombination would extend the harmonic cut-off due to the increase in ionization potential. 
To further estimate extension of HHG spectra one has to consider possible electronic excitations driven simultaneously to HHG process. In the present case, the excitation to LUMO and subsequent recombination would extend the spectrum by 1-2 more harmonics and we show the HOMO - LUMO gap and the values for the cut off extended in this way in the right two columns in Table \ref{tab:important_energies}.  Let us note that because of the approximations involved in these estimates, we do not show only integer values for the values of the cut-off harmonics. 

\begin{table}[h!]
    \centering
    \resizebox{1\columnwidth}{!}{\begin{tabular}[t]{c|cc|cc|cc}
        \hline
        \textbf{Molecule}  &  \textbf{Vertical $I_p$} & \textbf{Cut-off} & \textbf{Vertical $I_p$} & \textbf{Cut-off Harmonic} & \textbf{HOMO-LUMO gap} & \textbf{Maximum cut off} \\
        & (eV) & \textbf{Harmonic} & (with HOMO-1) (eV) & (with HOMO-1) & (eV)  & \textbf{extension} \\
        \hline
        \hline
        13CHD&  8.3099 & 9.8 & 11.6731 & 12.0 & 3.41 & 14.2\\
        13CHD*&  4.8397 & 7.6 & 8.4482 & 9.9 & 3.41 & 12.1\\
        cZcHT& 7.8256 & 9.5 & 10.0623 & 11.0 & 2.24 & 12.4\\
        cZtHT& 8.3861 & 9.9 & 10.4895 & 11.2 & 2.58 & 12.9\\
        135HT& 8.4297 & 9.9 & 10.3017 & 11.1 &  3.17 & 13.1 \\
        \hline
    \end{tabular}}
    \caption{Ionization potentials for isomers of interest and the corresponding approximations for cut-off harmonics. }
    \label{tab:important_energies}
\end{table}%
 
Since the amplitude and phases of the peaks in the harmonic spectrum are dependent on the electronic properties, geometry and symmetry of the electronic state of the system, looking at the HHG spectrum for each of the isomers can be used to indicate the presence of given isomer in the ensemble of the molecules and provide a method for the detection of the formation of each isomer during the course of the reaction. 

\begin{figure}
    \centering
    \includegraphics[width=0.6\columnwidth]{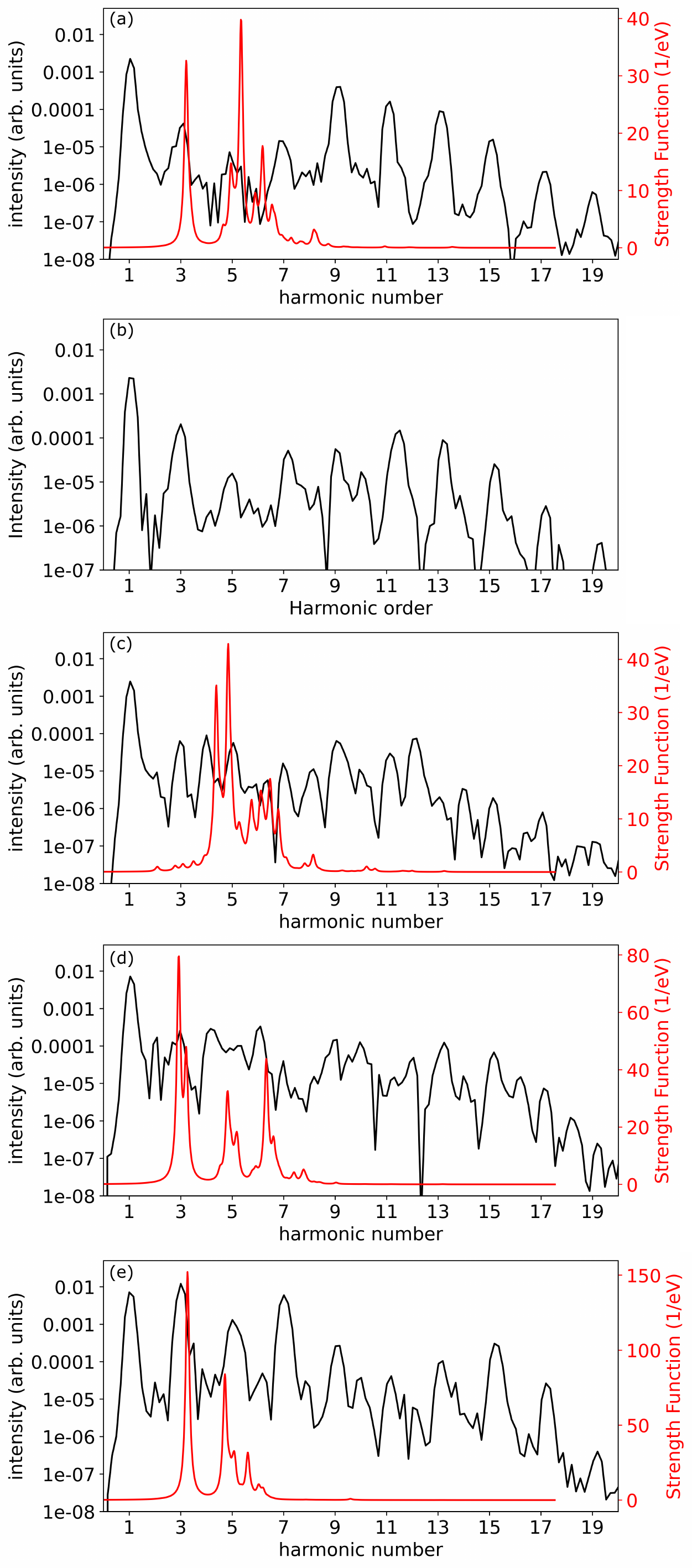}
    \caption{High harmonic spectra for (a) 1,3-cyclohexadiene, (b) excited 1,3-cyclohexadiene (c) cZc-1,3,5-hexatriene, (d) cZt-1,3,5-hexatriene, and (e) 1,3,5-(Z)-hexatriene for a $\hat{x}$-polarized laser pulse with a wavelength of 800 nm and an intensity of 3.7$\times$10$^{13}$W/cm$^{2}$. The Casida absorption spectrum is plotted in red.}
    \label{fig:hhg_x}
\end{figure}

Figure \ref{fig:hhg_x} shows results for HHG spectra for the 
1,3-cyclohexadiene (13CHD), cZc-1,3,5-hexatriene, cZt-1,3,5-hexatriene, and (d) 1,3,5-(Z)-hexatriene for $\hat{x}$ -polarized laser pulse with a
wavelength of 800 nm and an intensity of 3.7$\times$10$^{13}$W/cm$^{2}$.   The red solid line presents the results of the calculations of the one-photon absorption spectrum using the Casida method \cite{Casida1996,Casida1998} and peaks correspond to significant excitations parallel to the polarization of the laser field. 

Figure \ref{fig:hhg_x}(a) shows a result for the HHG spectrum for 13CHD.  One can notice that there are only odd harmonics likely due to 13CHD having "almost" C2 symmetry. Moreover, the spectrum exhibits suppression of the 3rd, 5th, and 7th harmonics. Because 13CHD is the reactant for the photoisomerization reaction, the spectrum can serve as a benchmark for detecting the changes related to the progress of the reaction. For example, any major increases in the intensities of these 3 harmonics would indicate that 13CHD is being consumed due to the reaction progress. Let us note that this suppression is also present for higher intensities, indicating that this property can be thought of as a characteristic signature of this particular molecule for this reaction.

When we compare the spectra for 13CHD and 13CHD$^*$ one can notice the 7th harmonic seems shifted and there is a minimum just before 9th harmonic. Because HOMO-LUMO transition energy is for this system at 11.293 eV which corresponds to 7.33$\omega_L$ we can expect the change between 7th and 9th harmonic to be related to the interference of the HHG process with the emission of the photon during the process of the deexcitation for 13CHD$^*$. Such a feature can be broader, in particular for shorter pulses with broader bandwidth, like it is in the present case, and in fact, it can also be accompanied by the "satellite" peaks.   

When comparing results for 13CHD to the other HHG spectra for the other systems considered, one can observe that the intensity of the 11th harmonic is much larger than for the other isomers. Hence in addition to an increase in the intensity of the suppressed harmonics, any decrease in intensity for the 11th harmonic could be used as an indicator of the progress of the reaction. 

The spectrum for the first intermediate, cZcHT, shown in Figure \ref{fig:hhg_x}(c) exhibits even harmonics. The presence of even harmonics is in part due to the bond breaking, but also due to breaking the symmetry of the molecule. This symmetry breaking causes more atoms to be situated away from the molecular plane. The appearance of the 4th, 8th, and 12th harmonics can be interpreted as an indication of the formation of cZcHT during the reaction progress. 
Compared to cZtHT, the other isomer with a 12th harmonic shown in \ref{fig:hhg_x}(d), cZcHT spectrum exhibits higher intensity of that harmonic. A decrease in the intensity of this harmonic could be indicative of the consumption of cZcHT in the reaction process.

The HHG spectrum for cZtHT has both odd and even integer multiples of the driving laser pulse frequency (see Figure \ref{fig:hhg_x}(d)). Here the appearance of 2nd, 6th, and 10th harmonics could be used as an indication of the presence of cZtHT, since these even harmonics are unique for this isomer. There is also major suppression of the 5th and 7th harmonics, so consequently any major decrease in the intensity of those harmonics could also serve as an indicator of the presence of cZtHT. 

When looking at the spectrum for the final product, 135HT, in Figure \ref{fig:hhg_x}(e) there is once again the presence of only odd harmonics. This is due in part to 135HT having a plane of symmetry. The intensities of the 3rd, 5th, and  7th harmonics are much higher than the corresponding harmonics for the other isomers, making any large increase in the intensities of those harmonics a marker for the formation of 135HT. In addition, another remarkable feature appearing for this isomer is a minimum for the 11th harmonic. 

All of these properties of the intensities of odd harmonics, as well as the presence of certain even harmonics, and relative suppression of the intensities for certain harmonics, can be used for the discrimination of isomers as well as for "timing" the reaction in a pump-probe experiment when HHG is used as a probing method.

The Casida equation is a pseudo-eigenvalue equation written in the basis of particle-hole states within many-body theory. In the calculations, one needs both the occupied states – computed during the ground-state calculation – as well as the unoccupied states. In the present case computed absorption spectra using the Casida method have 2-3 more prominent peaks with energies corresponding the region between the 3rd and 7th harmonic, as can be seen in Figure \ref{fig:hhg_x} and each peak is related to multiple transitions that in our calculations reach as high as LUMO+4. 
Moreover in the context of HHG radiation the results for Casida absorption spectra should be analyzed as a possible indication of rather weak contributions from excitations, because the absorption spectrum obtained with the Casida method, shown in Figure \ref{fig:hhg_x}, represents absorption due to one photon transitions and in the case of 800 nm wavelength of the laser pulse the transitions visible as peaks in the absorption spectrum would require correspondingly absorption of 3, 5, or 7 photons. HHG radiation intensity of 5-6 orders of magnitude lower than that of the driving laser. So we expect these excitations to not play a significant role and only present results as an indicator that in principle it seems the excitations can affect the regions of the HHG spectra corresponding to 3rd and 5th harmonics for 13CHD and 135HT,  5th and 7th for cZcHT and 3rd, 5th and 7th for cZtHT. One can notice that the HHG spectrum for 13CHD might have minimum affecting the 3rd, 5th, and 7th harmonics that might be related to interferences of direct HHG signal with HHG modified by excitations. 
From all of the excitations  included the excitations 
to LUMO from HOMO, HOMO-1, HOMO-2, HOMO-3 and HOMO-4
have the highest probability and excitations to LUMO+1 
from HOMO
are 10 to 100 times less probable so for example we did not include the LUMO+1 and any higher-lying orbitals in our estimates of the extension of the cut-off. 

Because the excitations are dependent on the isomer and they contribute to the specific characteristic properties of the HHG spectra discussion of the excitations does not change the conclusions regarding the discrimination of the isomers for the present study. 

In the next subsection, we present results for time dependent electron density and multiphoton ionization.

\subsection{Time-dependent electron density and electron density difference and ionization}
The electron density can be used to directly visualize changes in the system that could be related to the features shown in the HHG spectra but also, it is interesting to look at the real-time response to the ultrafast intense laser pulse. Direct simulations of the propagation for the molecular wavepacket on the grid allow us to peek inside the molecule and learn about features of the electron dynamics and the electronic properties of polyatomic molecules that we consider in this paper. Because HHG is a highly nonlinear nonperturbative process sometimes small changes in the electron dynamics in a molecule can lead to new features in the harmonic spectrum. In order to better visualize the changes due to interaction with the laser pulse the electron density difference is plotted in addition to the total electron density. The electron density difference is defined as $\rho(t,\textbf{r})-\rho(t=0,\textbf{r})$.
The biggest changes due to interaction with linearly polarized laser pulse are expected along the laser pulse polarization axis, $\hat{x}$, therefore we integrate over the other two dimensions in order to visualize the laser-driven electron dynamics. The integration is performed for the 2 dimensions over the corresponding sizes of the simulation box, using the standard formula,
\begin{equation}\label{integration}
   \rho(x,t) = \int_{y_{min}}^{y_{max}}\int_{z_{min}}^{z_{max}} \rho(x,y,z,t) dy dz
\end{equation}
where $y_{min}$, $y_{max}$, $z_{min}$, and $z_{max}$ are the limits of the simulation box. 
\begin{figure}
    \centering
    \includegraphics[width=0.7\columnwidth]{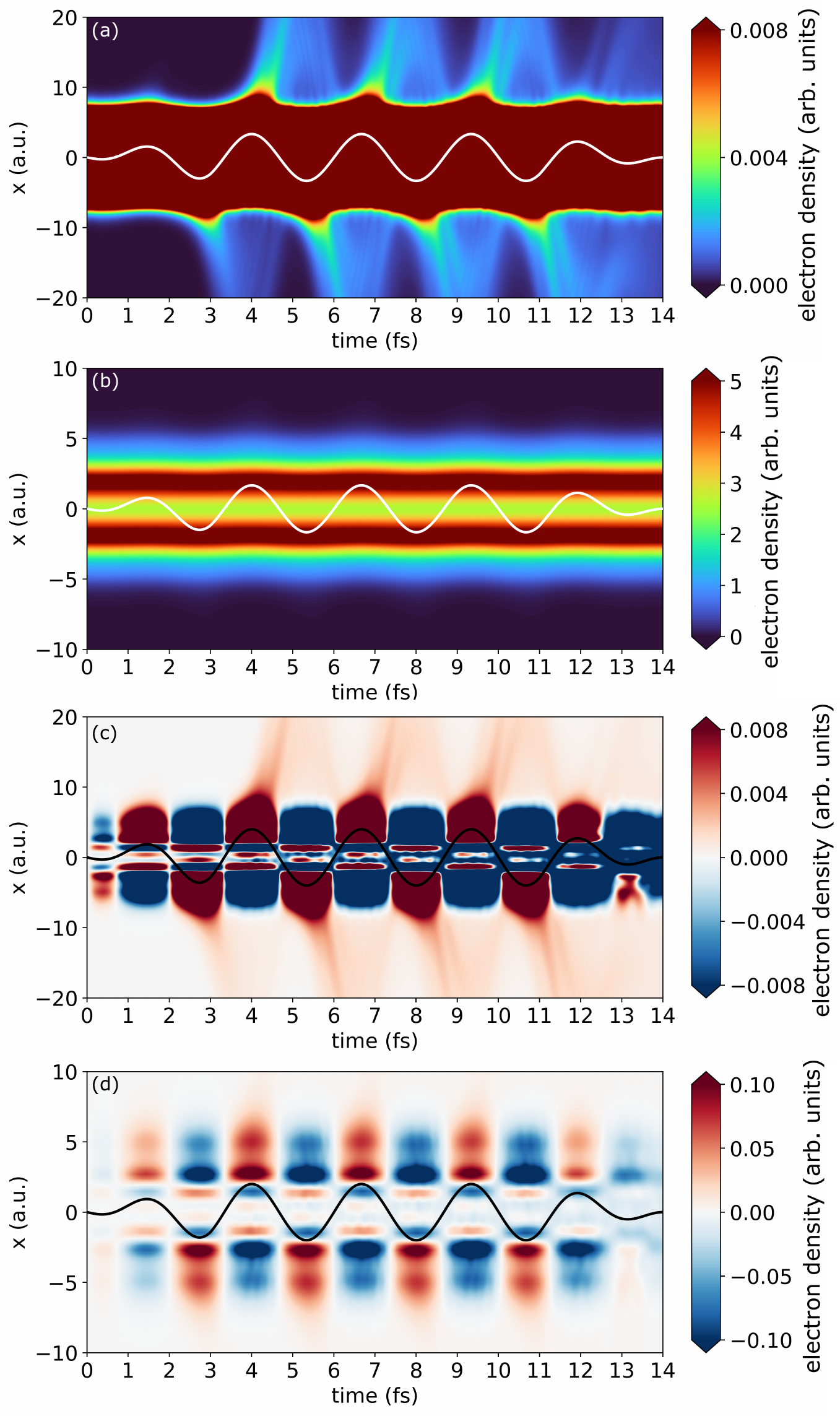}
    \caption{Integrated electron density for 1,3-cyclohexadiene plotted until (a) twice the quiver radius and (b) the molecular region and integrated electron density difference plotted until (c) twice the quiver radius and the (d) molecular region. The vector potential is plotted in white (a, b) or black (c, d).}
    \label{fig:13CHDdens}
\end{figure}

\begin{figure}
    \centering
    \includegraphics[width=0.7\columnwidth]{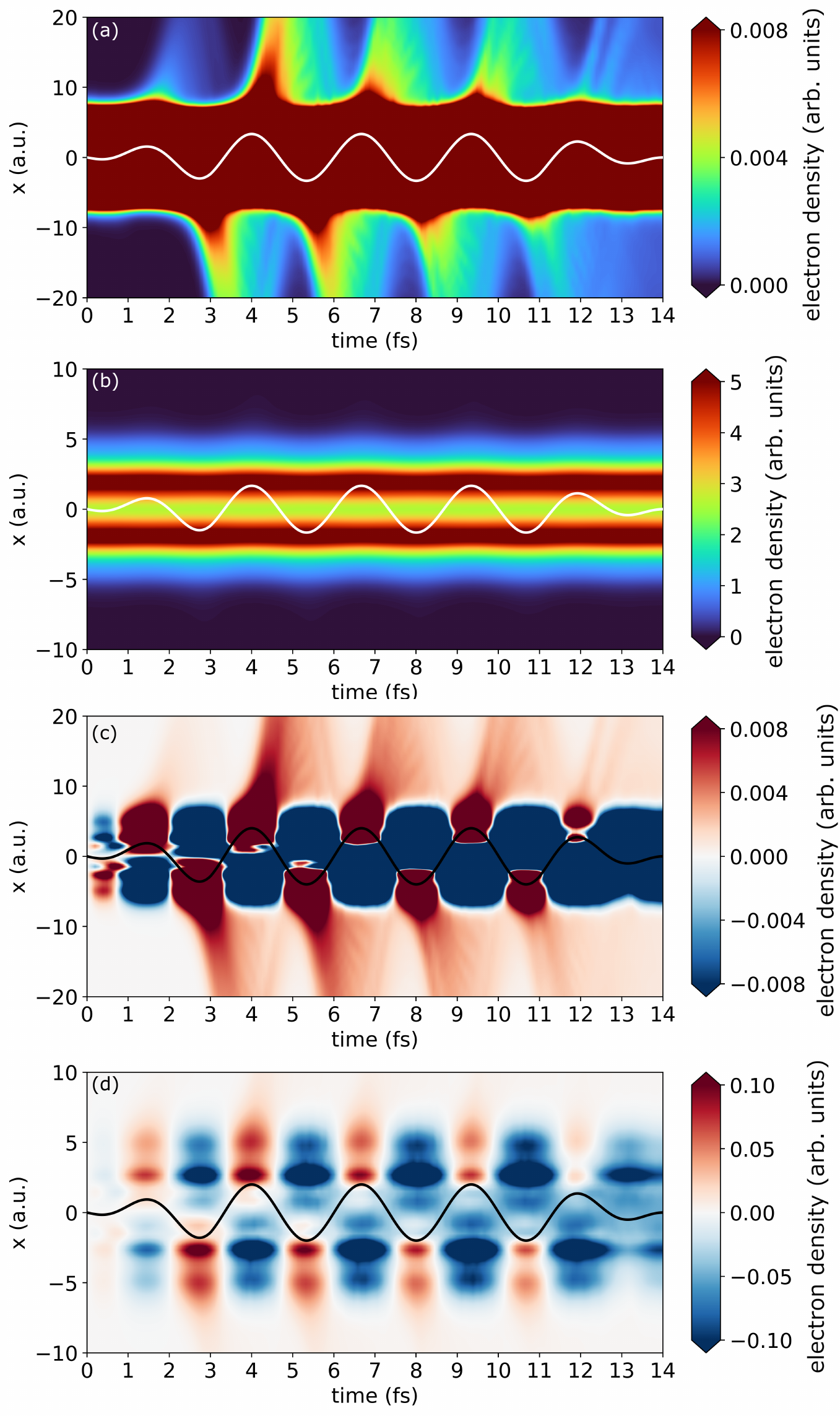}
    \caption{Integrated electron density for excited 1,3-cyclohexadiene plotted until (a) twice the quiver radius and (b) the molecular region and integrated electron density difference plotted until (c) twice the quiver radius and the (d) molecular region. The vector potential is plotted in white (a, b) or black (c, d).}
    \label{fig:13CHDexdens}
\end{figure}

\begin{figure}
    \centering
    \includegraphics[width=0.7\columnwidth]{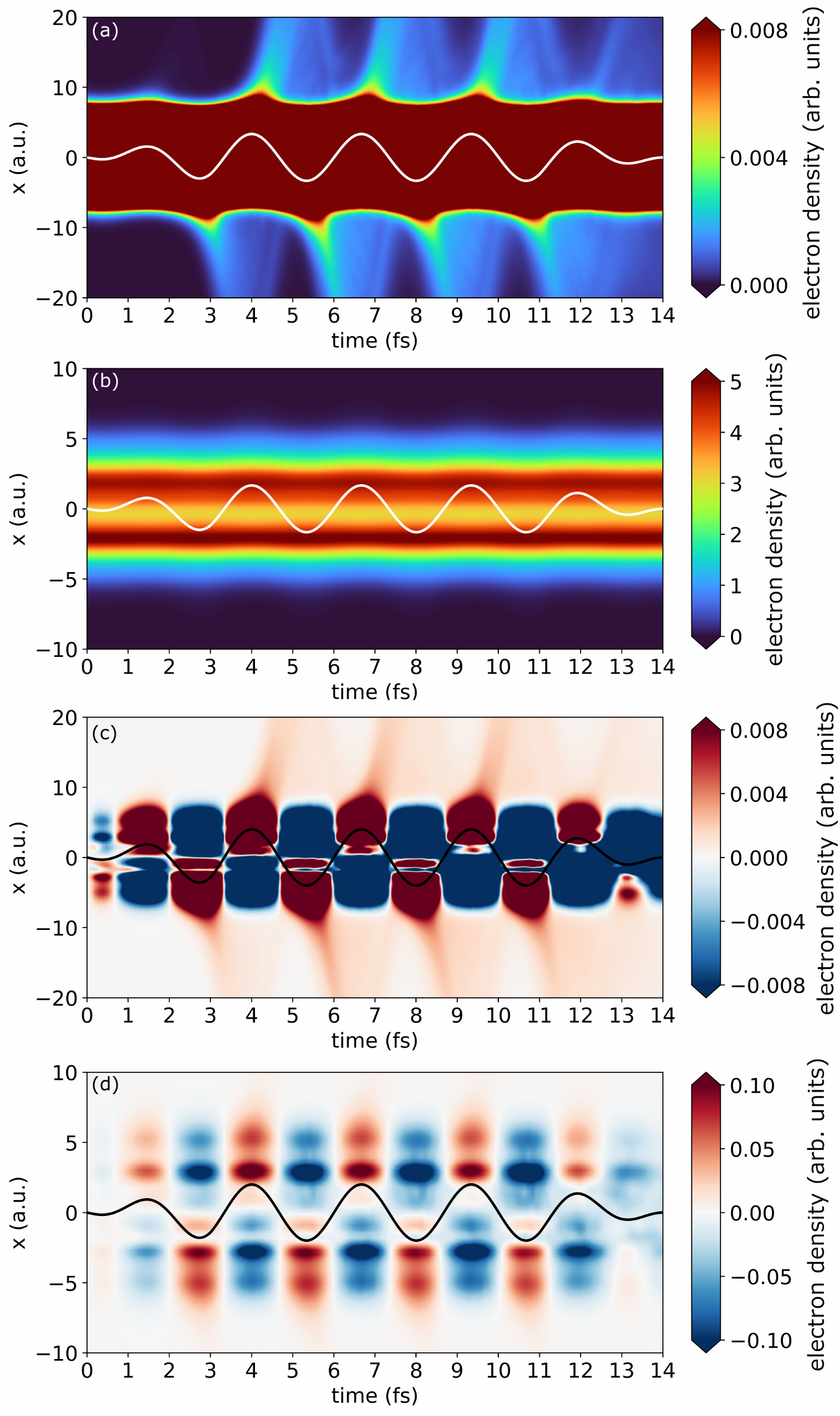}
    \caption{Integrated electron density for cZc-hexatriene plotted until (a) twice the quiver radius and (b) the molecular region and integrated electron density difference plotted until (c) twice the quiver radius and the (d) molecular region. The vector potential is plotted in white (a, b) or black (c, d).}
    \label{fig:cZcHTdens}
\end{figure}

\begin{figure}
    \centering
    \includegraphics[width=0.7\columnwidth]{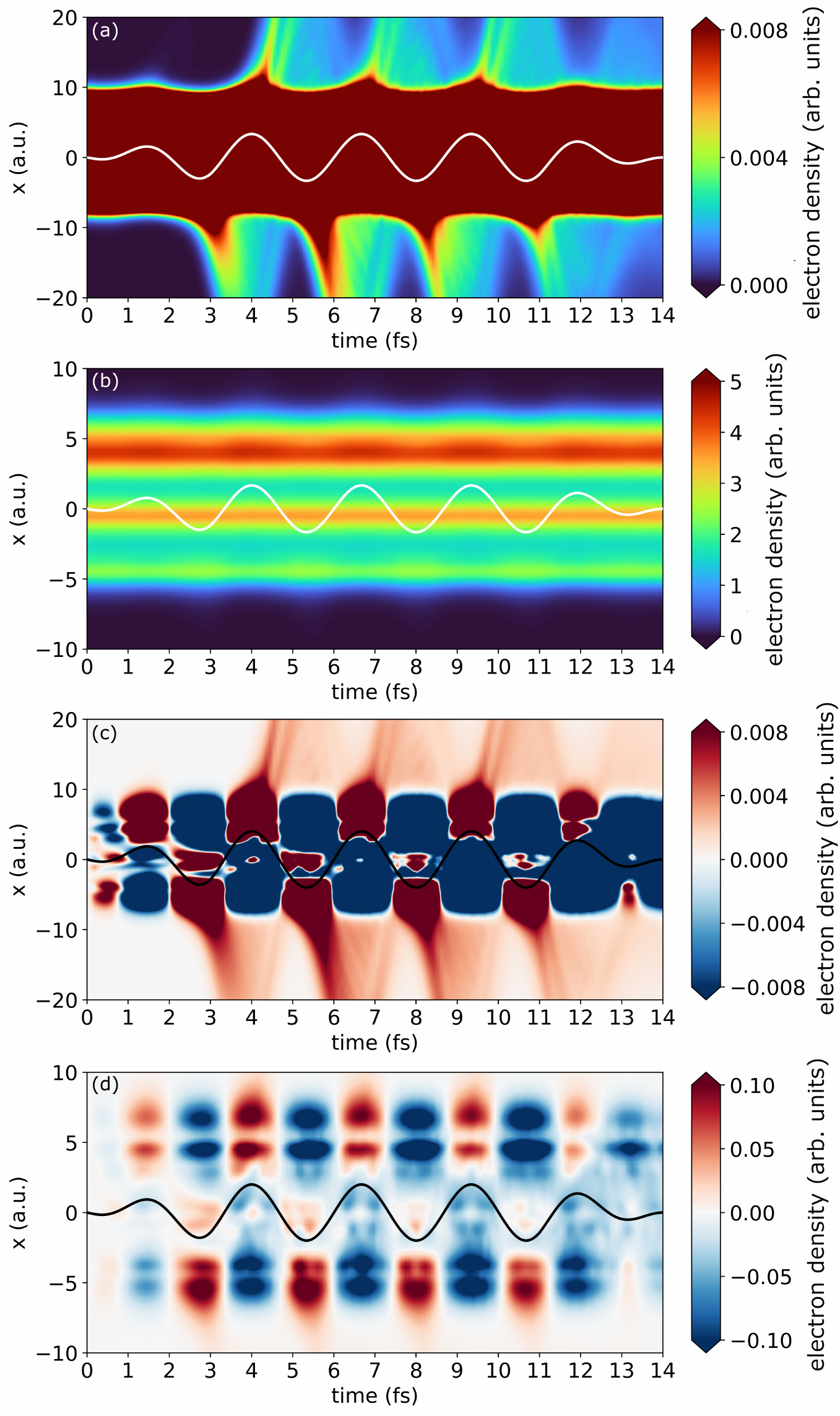}
    \caption{Integrated electron density for cZt-hexatriene plotted until (a) twice the quiver radius and (b) the molecular region and integrated electron density difference plotted until (c) twice the quiver radius and the (d) molecular region. The vector potential is plotted in white (a, b) or black (c, d).}
    \label{fig:cZtHTdens}
\end{figure}

\begin{figure}
    \centering
    \includegraphics[width=0.7\columnwidth]{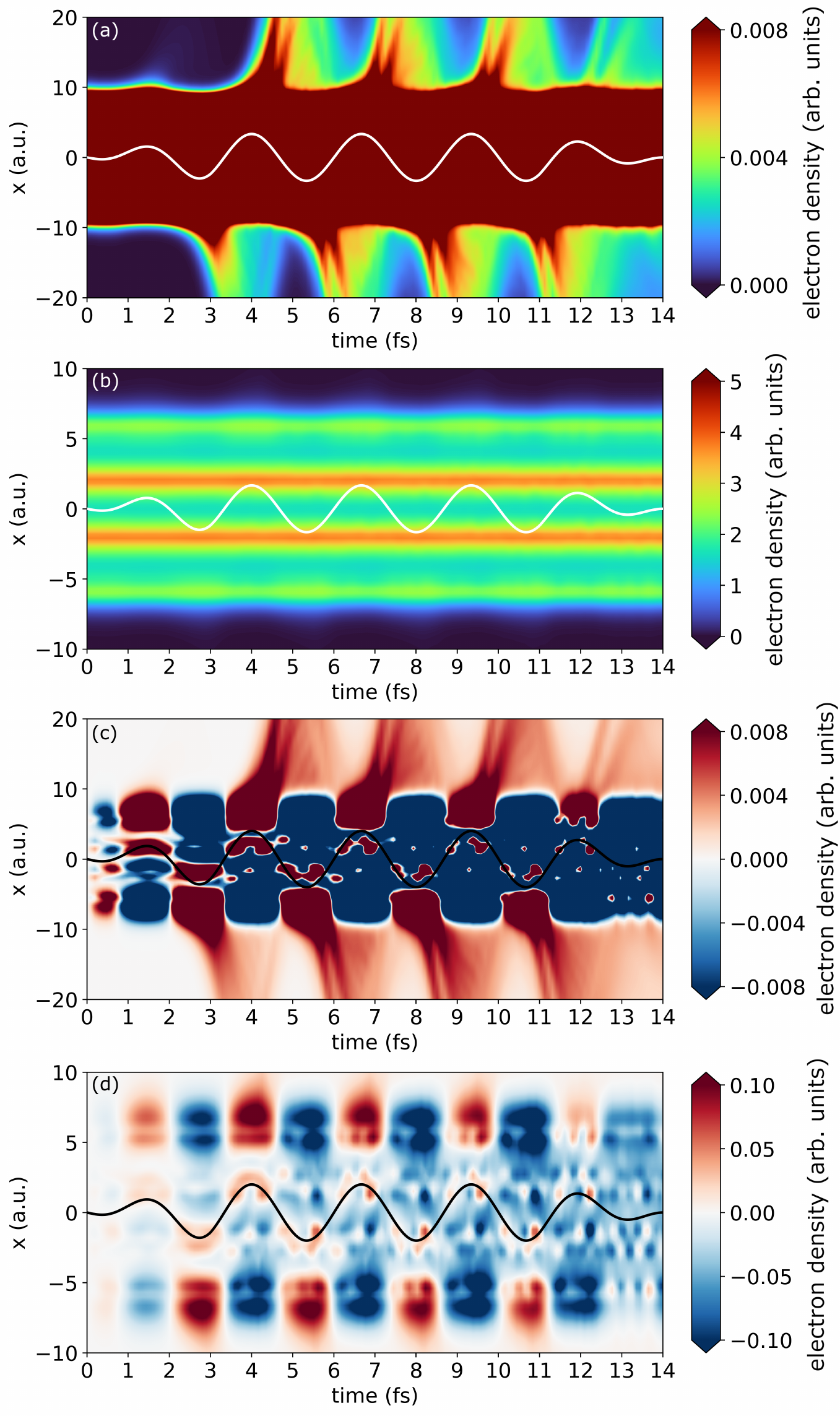}
    \caption{Integrated electron density for 1,3,5-(Z)-hexatriene plotted until (a) twice the quiver radius and (b) the molecular region and integrated electron density difference plotted until (c) twice the quiver radius and the (d) molecular region. The vector potential is plotted in white (a, b) or black (c, d).}
    \label{fig:135HTdens}
\end{figure}

\begin{table}[h!]
    \centering
    \begin{tabular}[t]{l|cc}
        \hline
        \textbf{Molecule}  & \textbf{Total Ionization (a.u.)} \\
        \hline
        \hline
        13CHD & 0.1730 \\
        13CHD* & 0.5549 \\
        cZcHT & 0.1963 \\
        cZtHT & 0.3879 \\
        135HT & 0.5797 \\
    \end{tabular}
    \caption{Total ionization at 14 fs (end of laser pulse) for all molecules calculated.}
    \label{tab:ionization}
\end{table}

The three-step model for harmonic generation can further support analysis of the integrated electron density and electron density difference shown as a function of time in Figures \ref{fig:13CHDdens}(a) to \ref{fig:135HTdens}(a). First one notices that at times corresponding to the maximum/minimum of the electric field the outgoing wavepacket can be seen starting the excursion toward the end of the grid.  
For each molecule and for each electron density and electron density difference we show one figure for density and density difference shown up to twice the quiver radius with the rescattering region included as well as the zoomed part to show changes of the electron density and electron density difference in the "molecular region". 
Let us note that we plot the results for all isomers on the same scale so that we can compare the corresponding changes in color intensity for a given feature in the plot. 
Since we present results of quantum mechanical simulations, integrated over 2 spatial dimensions and for example, changes due to HHG are expected to happen at the scale of 10$^{-6}$ lower than the main part of the wavepacket, we do not expect to see clear signatures that can be uniquely matched to 
rescattering and recombination events or even classical trajectories in our plots. However one can see the features like returning wavepacket and the spreading of the outgoing wavepacket.  

We will now attempt the qualitative description of the features in the plots. 
The closest to visualization of the "rescattering wavepacket" in the plots is the portion of the wavepacket indicated by the deeper red color at the electron density difference plots, 
Figures \ref{fig:13CHDdens}(c) to \ref{fig:135HTdens}(c). We will focus first on the comparison of the properties of this feature. Changes in this feature are partially due to the fact that atoms in the molecule are located closer to the $\hat{x}$-axis as the reaction proceeds.   
For 13CHD* represents a special case and has an outgoing wavepacket intensity that falls closer to being between that of cZtHT and 135HT. This is likely due to the molecule being in an excited state,  which decreases both the vertical and adiabatic ionization potential energy,  as shown in Table \ref{tab:important_energies}.
The trend of wavepacket intensity having local maxima for 13CHD* when looking at isomerization progress can also be seen in Table \ref{tab:ionization} which shows total ionization at the end of the laser pulse. This ionization is taken to be the portion of the outgoing wavepacket that is absorbed by the complex absorbing potential which represents the portion of an ensemble of molecules that is ionized.

For Figures \ref{fig:13CHDdens}(b) to \ref{fig:135HTdens}(b) the changes to the geometry and conjugation can be seen throughout the duration of the laser pulse. As the reaction proceeds, there are more carbon centers along the $\hat{x}$-direction which is causing the increase in density across the "molecular" region within the quiver radius. The changes in conjugation can also be seen as there is a more even spread of electron density for cZtHT and 135HT, shown in Figures \ref{fig:cZtHTdens}(b) and \ref{fig:135HTdens}(b). 
Moreover all of the isomers have the largest changes to the electron density near the atomic centers in the polarization direction as seen in Figures \ref{fig:13CHDdens}(d) to \ref{fig:135HTdens}(d). For 13CHD and cZcHT there are minimal changes to the density in the bonding region between atomic centers as seen in  Figures \ref{fig:13CHDdens}(d) and \ref{fig:cZcHTdens}(d),  likely due to the lack of full conjugation of the molecules. 

For 13CHD*, cZtHT, and 135HT there is suppression in the electron density between the atoms, in the bonding region, after the onset of ionization (around 3 fs) as shown in Figures \ref{fig:13CHDexdens}(d), \ref{fig:cZtHTdens}(d),  and \ref{fig:135HTdens}(d). For 13CHD* this decrease in density is likely due to the wavepacket representing excited system and exhibiting enhanced ionization. This results in changed spatial extension, modified bonding, and changes in density values across the molecule.

Most changes to 13CHD, 13CHD*, cZcHT occur around the atoms in the laser pulse polarization direction as seen in Figures \ref{fig:13CHDdens}(d), \ref{fig:13CHDexdens}(d), and \ref{fig:cZcHTdens}(d). Most dramatic changes for cZtHT and 135HT in electron density occur between the atoms, for $-5 a.u. < x < 5 a.u.$, as can be seen in Figures \ref{fig:cZtHTdens}(d) and \ref{fig:135HTdens}(d), where the changes in electron density are not as smooth and regular like the other isomers discussed. These qualitative differences in electron dynamics is likely due to cZtHT and 135HT not only being fully conjugated but also planar. The effect of conjugation can also be seen in Figures \ref{fig:cZtHTdens}(b) and \ref{fig:135HTdens}(b) where the electron density is spread evenly across the $\hat{x}$-axis.

\section{Summary}
We present the results of a theoretical study inspired by a more recent experiment designed to follow a photoisomerization reaction by probing it with high-order harmonic generation process. The reaction is initiated by the UV pump pulse. We assume that the reaction further proceeds without the laser field until at a given time delay the intense probe laser pulse is used to generate high order harmonics. We conclude that HHG spectra have characteristic features for each of the system considered. Harmonic intensity  and total ionization are all possible candidates for observables enabling real-time tracking of the reaction progress.  

Moreover, we present the visualization of the interaction of the isomers with ultrashort intense laser pulses for integrated electron density and integrated electron density difference along laser pulse polarization direction, as a function of time. The changes in the structure of the isomers can be seen influencing changes in the molecular and rescattering regions as presented in the results. 

\section{Acknowledgements}
The authors acknowledge support from National Science Foundation PHY-2317149 and PHY-2110628 awards. 
Our work utilized the Alpine high-performance computing resource at the University of Colorado Boulder.
Alpine is jointly funded by the University of Colorado Boulder, the University of Colorado Anschutz, and Colorado State University. This work also utilized the Summit supercomputer, which is supported by the National Science Foundation (awards ACI-1532235 and ACI-1532236), the University of Colorado Boulder, and Colorado State University. 

\bibliography{apssamp}

\end{document}